\documentclass[reprint,aps,pra,amsmath,amssymb,superscriptaddress,floatfix,longbibliography]{revtex4-2}
\usepackage{amstext}
\usepackage{graphicx}

\begin{document}
\title{Reaching Quantum Critical Point by Adding Non-magnetic Disorder\\ in Single Crystals of (Ca$_{x}$Sr$_{1-x}$)$_{3}$Rh$_{4}$Sn$_{13}$ Superconductor}

\author{Elizabeth~H.~Krenkel}
\affiliation{Ames National Laboratory, Ames, Iowa 50011, U.S.A.}
\affiliation{Department of Physics \& Astronomy, Iowa State University, Ames, Iowa
50011, U.S.A.}

\author{Makariy~A.~Tanatar}
\affiliation{Ames National Laboratory, Ames, Iowa 50011, U.S.A.}
\affiliation{Department of Physics \& Astronomy, Iowa State University, Ames, Iowa 50011, U.S.A.}

\author{Romain~Grasset}
\affiliation{Laboratoire des Solides Irradi\'{e}s, CEA/DRF/lRAMIS, \'{E}cole Polytechnique, CNRS, Institut Polytechnique de Paris, F-91128 Palaiseau, France}

\author{Marcin~Ko\'{n}czykowski}
\affiliation{Laboratoire des Solides Irradi\'{e}s, CEA/DRF/lRAMIS, \'{E}cole Polytechnique, CNRS, Institut Polytechnique de Paris, F-91128 Palaiseau, France}

\author{Shuzhang~Chen}
\affiliation{Condensed Matter Physics and Materials Science Department, Brookhaven
National Laboratory, Upton, New York 11973, U.S.A.}
\affiliation{Department of Physics and Astronomy, Stony Brook University, Stony
Brook, New York 11794-3800, U.S.A.}

\author{Cedomir~Petrovic}
\affiliation{Condensed Matter Physics and Materials Science Department, Brookhaven
National Laboratory, Upton, New York 11973, U.S.A.}
\affiliation{Department of Physics and Astronomy, Stony Brook University, Stony
Brook, New York 11794-3800, U.S.A.}
\affiliation{Shanghai Key Laboratory of Material Frontiers Research in Extreme Environments (MFree), Shanghai Advanced Research in Physical Sciences (SHARPS), Pudong, Shanghai 201203, China}
\affiliation{Department of Nuclear and Plasma Physics, Vinca Institute of Nuclear Sciences, University of Belgrade, Belgrade 11001, Serbia}

\author{Alex~Levchenko}
\affiliation{Department of Physics, University of Wisconsin--Madison, Madison,
Wisconsin 53706, U.S.A.}

\author{Ruslan~Prozorov}
\affiliation{Ames National Laboratory, Ames, Iowa 50011, U.S.A.}
\affiliation{Department of Physics \& Astronomy, Iowa State University, Ames, Iowa 50011, U.S.A.}

\date{21 June 2024}

\begin{abstract}
The quasi-skutterudites (Ca$_{x}$Sr$_{1-x}$)$_{3}$(Rh, Ir)$_{4}$Sn$_{13}$ show a rare nonmagnetic quantum critical point associated with the second-order charge-density-wave (CDW) and structural distortion transition extended under the superconducting ``dome''. So far, the non-thermal tuning parameters for accessing the QCP included changing stoichiometry, pressure, and a magnetic field. Here we add another parameter -- a nonmagnetic point-like disorder induced by 2.5 MeV electron irradiation. The non-Fermi liquid regime was inferred from the analysis of the temperature-dependent resistivity, $\rho\left(T\right)$, in single crystals of (Ca$_{x}$Sr$_{1-x}$)$_{3}$Rh$_{4}$Sn$_{13}$. Starting at compositions below the known QCP concentration of $x_c=0.9$, added disorder resulted in a progressively larger linear term and a reduced quadratic term in $\rho\left(T\right)$. This behavior is supported by theoretical analysis based on the idea of superconducting fluctuations encompassing the crossover from quantum to thermal regimes. Our results strongly support the concept that the nonmagnetic disorder can drive the system toward the quantum critical regime.
\end{abstract}

\maketitle

\section*{Introduction}

Among the quasi-skutterudite Remeika 3-4-13 compounds, Ca$_{3}$Rh$_{4}$Sn$_{13}$ has one of the highest superconducting transition temperatures, at $T_c=8.3\:\text{K}$ \cite{Remeika1980}. It does not show a charge-density wave (CDW) or the associated structural distortion. In contrast, its sibling, Sr$_{3}$Rh$_{4}$Sn$_{13}$, develops CDW below $T_{\text{CDW}}=136\:\text{K}$ and superconductivity below $T_{c}=4.6\:\text{K}$.  Naturally, alloy compositions, (Ca$_{x}$Sr$_{1-x}$)$_{3}$Rh$_{4}$Sn$_{13}$, have been investigated for a possible quantum phase transition (QPT), which was found at ambient pressure at the critical concentration, $x_{c}=0.9$, and also by varying isostatic pressure, $P$ \cite{Klintberg2012,Goh2015,Teraski2021}. X-ray measurements confirmed that the $T_{\text{CDW}}\left(x,P\right)$ exists below the superconducting dome \cite{Veiga2020,Carneiro2020}.  A variety of properties exhibit singular behavior at this structural/CDW quantum critical point (QCP).  In addition to ``classical'' non-Fermi behavior from the temperature dependence of the resistivity above $T_{c}$, the specific heat jump at the transition is larger than a weak-coupling value. This is in line with the enhanced superfluid density and pairing strength at the QCP determined by muon spin rotation spectroscopy \cite{Krieger2023}.  The critical current also shows a sharp peak at the QCP composition at temperatures which are deep inside the superconducting dome, further confirming its existence \cite{Liu2022}. In another 'sibling' compound, the refractory metal site is instead permuted, bounded by Ca$_{3}$Rh$_{4}$Sn$_{13}$ (no CDW and $T_{c}=8.3\:\text{K}$) and Ca$_{3}$Ir$_{4}$Sn$_{13}$ ($T_{\text{CDW}}=39\:\text{K}$ and $T_{c}=6.9\:\text{K}$). Recent transport and magnetization measurements indicate that in Ca$_{3}$(Rh$_{x}$Ir$_{1-x}$)$_{4}$Sn$_{13}$ the CDW is suppressed somewhere between $x=0.53$ and $0.58$, and the temperature-dependent resistivity deviates from the Fermi liquid behavior, perhaps implying a QCP in this region \cite{Krenkel2023}.

\begin{figure}[tb!]
\includegraphics[width=8cm]{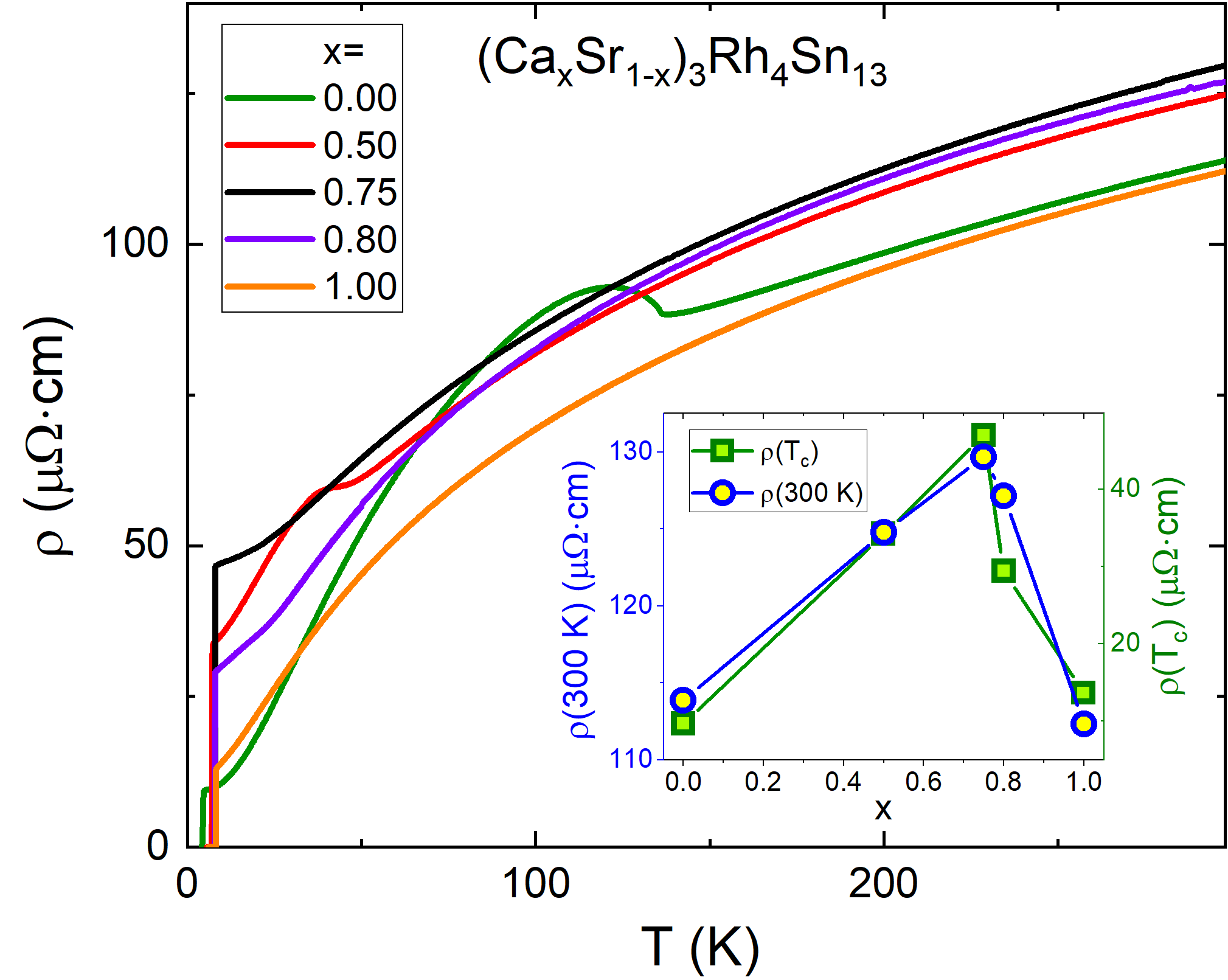}
\caption{Temperature-dependent resistivity, $\rho\left(T\right)$,
of pristine samples of (Ca$_{x}$Sr$_{1-x}$)$_{3}$Rh$_{4}$Sn$_{13}$.
Inset shows compositional evolution of resistivity at two characteristic
temperatures, $\rho\left(T=300\;\text{K}\right)$ (left axis), and
at the onset of the superconducting transition, $\rho\left(T_c\right)$
(right axis). }
\label{fig:rho(T)} 
\end{figure}

To date, three nonthermal parameters have been used to tune various systems to a quantum critical point: composition, pressure, and magnetic field. However, it has been suggested that disorder might also serve as such a control parameter, particularly in cases where the long-range ordered-to-disordered quantum phase transition is not perfectly of the second order \cite{Belitz2005}. In these cases, disorder may push the transition toward a pure second-order transition and tune the system toward the QCP \cite{Belitz2005}. To our knowledge, there have been no reports of a controlled disorder-tuned QCP in any material or for any type of long-range order. We note that while chemical substitution influences the chemical potential level via charge doping \cite{Joshi2020} and the electronic band structure via chemical pressure \cite{Hashimoto2012}, it also introduces additional scattering, which was taken into account in the theoretical analysis of QCP \cite{Joshi2020}. However, it is impossible to separate these effects and isolate only the effects of scattering.

In this paper, we investigate the effects of disorder on coexisting CDW and superconductivity and demonstrate that the QCP can indeed be reached, by introducing a non-magnetic point-like disorder through 2.5 MeV electron irradiation in (Ca$_x$Sr$_{1-x}$)$_3$Rh$_4$Sn$_{13}$ quasi-skutterudite compounds starting with compositions well below the ``compositional" QCP located at $x_{c}=0.9$.


\section*{Results}

Figure \ref{fig:rho(T)} shows the temperature-dependent resistivity, $\rho\left(T\right)$, of several compositions studied of pristine samples of (Ca$_{x}$Sr$_{1-x}$)$_{3}$Rh$_{4}$Sn$_{13}$, indicated in the legend. The inset shows the compositional evolution of $\rho\left(T\right)$ at two characteristic temperatures, $\rho\left(T=300\;\text{K}\right)$ (left $y-$axis), and at the onset of the superconducting transition, $\rho\left(T_c\right)$ (right $y-$axis).  To determine the absolute value of the resistivity, we use the geometric factor generated by matching the slope of each curve at high temperature to the known slope of the resistivity of Ca$_3$Rh$_4$Sn$_{13}$ \cite{Krenkel2022}.  As expected, the stoichiometric compositions, $x=0$ and $x=1$, show the lowest values because in mixed compositions, the resistivity increases due to substitutional disorder. This makes the initial starting scattering rate of the pristine state $x-$dependent and not well defined. Therefore, we should rely on the changes in the properties introduced by repeated irradiation runs.

\begin{figure}[tb!]
\includegraphics[width=8.6cm]{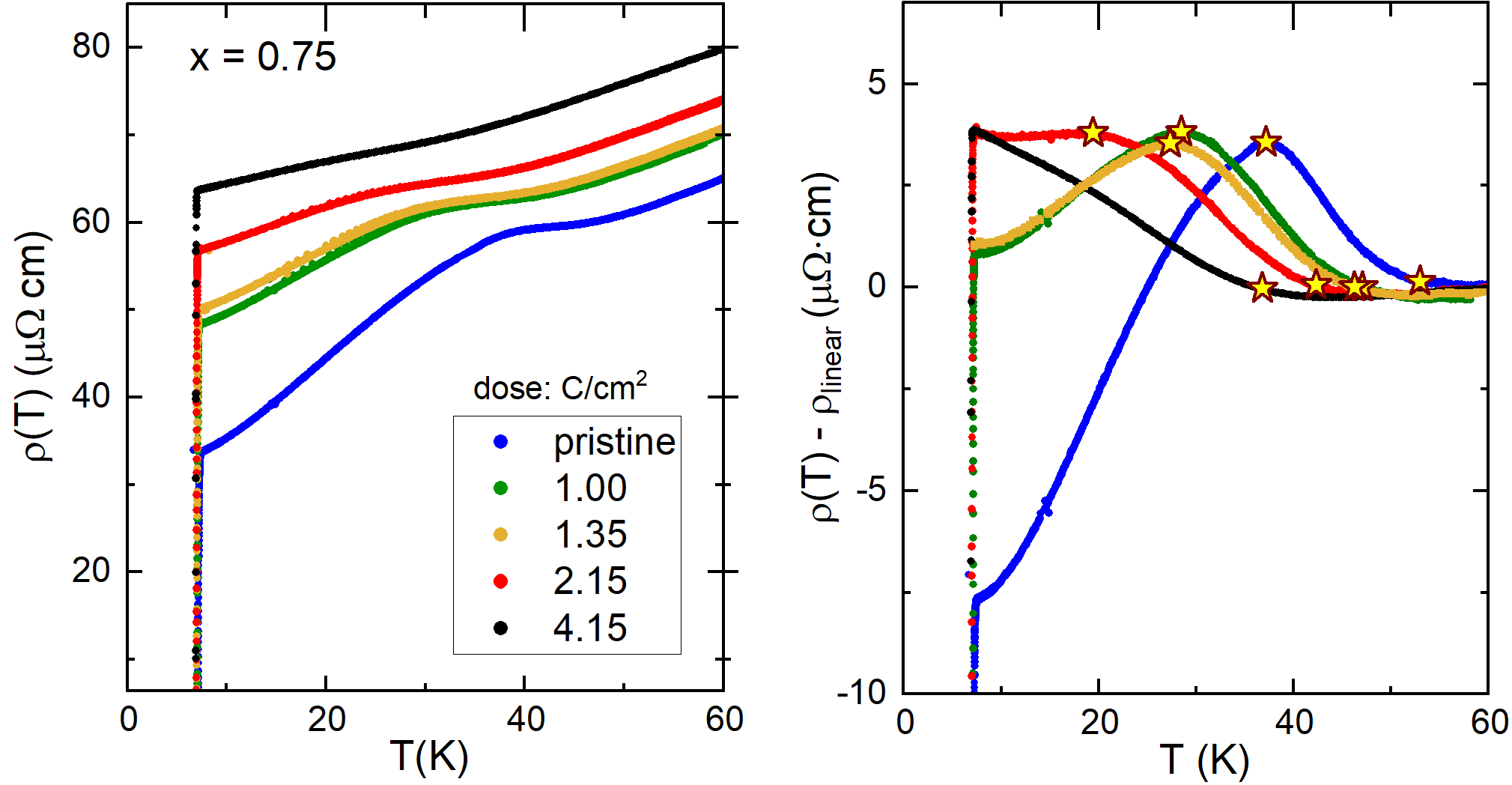}
\caption{Left panel: Measurements of the temperature-dependent resistivity, $\rho (T)$ with repeated electon irratiation, for the (Ca$_{0.75}$Sr$_{0.25}$)$_3$Rh$_4$Sn$_{13}$ composition.  The legend shows the accumulated irradiation doses. (Right panel):  Method of determination of the charge density wave regions of the material, via subtraction of a linear fit of the resistivity above the transition temperature.  Also clearly displays the crossover from predominately quadratic behavior (blue pristine curve) to predominantly linear resistivity(black curve). }
\label{fig:rho(T)diffDose} 
\end{figure}

Figure \ref{fig:rho(T)diffDose} shows the evolution of temperature-dependent resistivity, $\rho\left(T\right)$, after irradiation in $x=0.5$ and $x=0.75$, close to the QCP composition, $x_{c}=0.9$. The sample with $x=0.75$ is the main focus of this study.  The suppression of CDW by disorder is evident. Not only does the transition temperature shift, but the transition itself becomes broader and less obvious. Similar behavior is observed in previously studied stoichiometric Sr$_{3}$Rh$_{4}$Sn$_{13}$ and Sr$_{3}$Ir$_{4}$Sn$_{13}$ \cite{Krenkel2022}. In alloyed compositions, the CDW features are significantly less sharp compared to stoichiometric ones \cite{Krenkel2023}. 

\begin{figure*}[t]
\includegraphics[width=17cm]{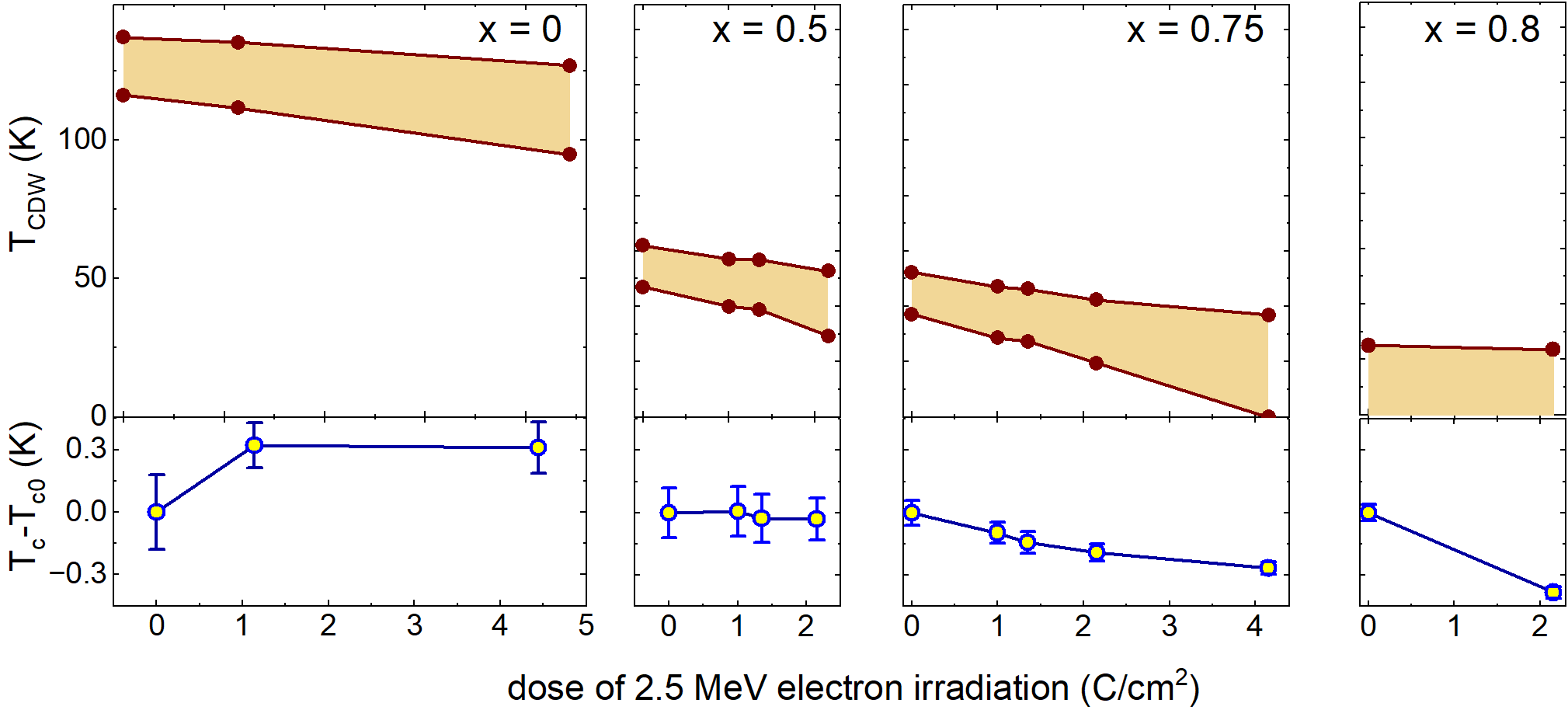}
\caption{(Color online) The evolution of the characteristic temperatures (Ca$_{x}$Sr$_{1-x}$)$_{3}$Rh$_{4}$Sn$_{13}$ with point-like disorder introduced by electron irradiation. (Top panels): the charge density wave transition, $T_{\text{CDW}}\left(x\right)$. The uncertainty in determining $T_{\text{CDW}}$, shown as shaded areas, is discussed in the text. (Bottom panels): superconducting transition temperature, $T_{c}\left(x\right)$. Due to a small variation, the latter is shown as the deviation from the initial $T_{c0}$ of the pristine samples. Note an initial increase of $T_{c}$ after irradiation in stoichiometric Sr$_{3}$Rh$_{4}$Sn$_{13}$.}
\label{fig:TcCDW-Dose} 
\end{figure*}

The T$_{\text{CDW}}$ was determined by subtracting a linear fit of the data above the transition temperature and then taking the point where the data began to deviate from that line, as the upper bound, and the peak of the difference between the two as the lower bound.  This procedure is shown in figure \ref{fig:rho(T)diffDose}, and the results are shown in Fig.\ref{fig:TcCDW-Dose} for samples of the compositions indicated in each panel. The top set of panels shows the evolution of the CDW transition, with a shaded area marking the temperature range between onset and offset, as discussed above. 

The bottom set of panels shows the evolution of the change in superconducting transition temperature, $\Delta T_{c}=T_{c}-T_{c0}$, where $T_{c0}$ is the transition temperature before irradiation. Interestingly, the stoichiometric Sr$_{3}$Rh$_{4}$Sn$_{13}$ shows initial increase of $T_{c}$ demonstrating the intrinsic interplay between the CDW and superconductivity. In the simplest scenario, the CDW gapping of the Fermi surface is weakened by the disorder, thereby enhancing superconductivity. Other mechanisms such as softening of the phonon modes are also possible, as observed in NbSe$_2$ \cite{KyuilNbSe2}. For the other compositions, the superconducting transition is suppressed, which is consistent with our previous conclusion of the unconventional pairing in stannide compounds \cite{Krenkel2022}.  However, the rate of change of $T_{c}$ in alloys is difficult to compare directly with the change of $T_c$ in the stoichiometric material. As mentioned above, there is an elevated substitutional disorder in the pristine state of the intermediate compositions, see Fig.\ref{fig:rho(T)}, and the corresponding discussion. Since in the absence of nodes, $T_{c}\left(\Gamma\right)$ is a saturating function of the dimensionless scattering rate, $\Gamma$, the rate of change, $dT_{c}\left(\Gamma\right)/d\Gamma$, is also a decreasing function of $\Gamma$.

\begin{figure*}[t]
\includegraphics[width=\textwidth]{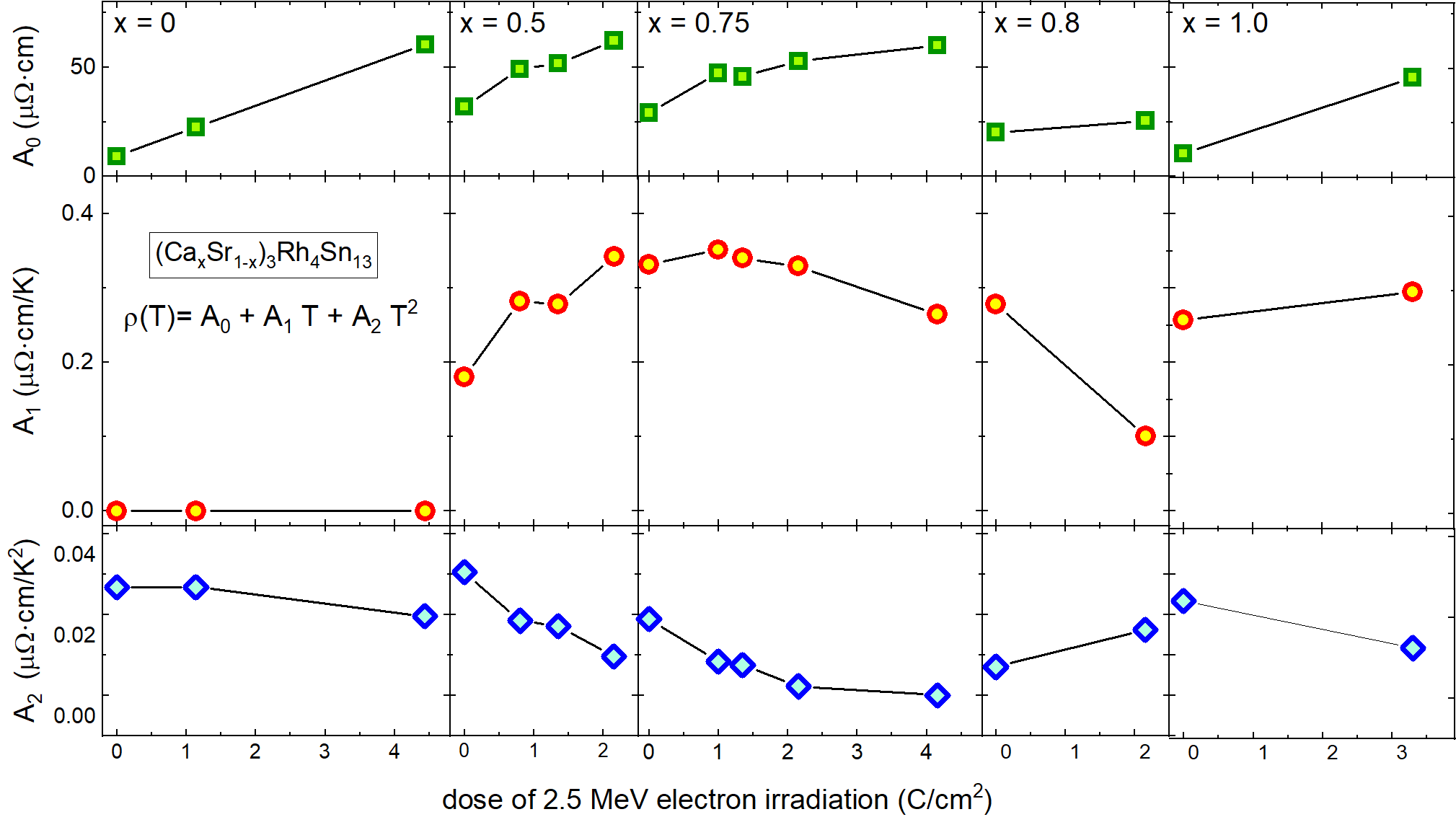}
\caption{(Color online) Summary of the doping/disorder evolution of the parameters of the polynomial fit $\rho=A_{0}+A_{1}T+A_{2}T^{2}$ of resistivity in the temperature range from 20 K to $T_{c}$. Panels from left to right $x=$0, $x=$0.5, $x=$0.75 and $x=$0.8, and $x=$1.}
\label{fig:polynomial} 
\end{figure*}

We now analyze the temperature-dependent resistivity above the transition temperature, $T_{c}$. As can be directly seen in Fig.\ref{fig:rho(T)diffDose},
$\rho\left(T\right)$, is nonlinear in the pristine state and is
practically linear at higher levels of disorder. We therefore performed a polynomial fitting of the resistivity curves, $\rho\left(T\right)=A_{0}+A_{1}T+A_{2}T^{2}$, for all compositions from above the transition temperature to 20 K. Our goal is to determine the evolution of the coefficients $A_{0,1,2}$ with disorder.  The result is summarized in Fig.\ref{fig:polynomial} for five different compositions. Note that each set of horizontal panels has a common vertical and horizontal scale to facilitate visual comparison.

The residual resistivity, $A_{0}$, increases for all samples as expected, indicating that the disorder was indeed added in each case. The most striking result is presented in the middle set of panels, where the linear coefficient indicative of non-Fermi liquid behavior is plotted. Although no linear term is detected for $x=0$, it becomes progressively more prominent with increasing disorder for $x=0.5$.  In $x=0.75$ initial increase is followed by a decrease.  In $x=0.8$, we only see a decrease.  Our interpretation is that this is because we overshot the QCP in these two compositions. Note that in these cases, we are adding disorder to already significant substitutional disorder level. 

Finally, the quadratic term, $A_{2}$, behaves in a way that is reciprocal to $A_{1}$. It drops significantly for intermediate compositions and remains practically unchanged for larger values. 

Overall, Fig.\ref{fig:polynomial} shows a proliferation of the linear temperature behavior of $\rho\left(T\right)$ with added non-magnetic disorder in the compositions not too far from QCP on the CDW ordered side of the compositional phase diagram. We interpret this behavior as non-magnetic disorder scattering acting as a nonthermal variable driving the system towards the quantum critical point, and through it for larger scattering rates.

\section*{Discussion}

We now provide general theoretical arguments and estimates for the evolution of $T$-linear and $T$-quadratic terms in $\rho(T)$ above $T_c$. Our analysis is limited to the parameter range close to the onset of superconductivity near $T_c(x)$, and, as such, it cannot quantitatively explain the data across the entire range of observed behavior at $T\gg T_c$. Nevertheless, it offers useful insights into the relevant physics, defining the possible mechanisms behind the observed features in the temperature and doping dependence of the resistivity, in particular near the critical point.

The idea we explore is based on the picture of superconducting fluctuations encompassing the crossover from quantum to thermal regimes, which are defined by proximity to the quantum critical point in the $T-x$ plane. Since our analysis pertains to temperatures $T>T_c(x)$, a microscopic model that captures the delicate interplay of superconducting (SC) and charge-density wave (CDW) orders beneath the dome will not be necessary. However, such a model in principle can be constructed based on earlier works, see e.g. Ref. \cite{Machida:JPSJ84}, with the necessary generalization to include disorder.

Near the critical temperature, the dependence of resistivity $\rho(T)$ is strongly affected by superconducting fluctuations \cite{Book}. The corresponding transport theory is well established for the classical regime of fluctuations, which can be formally defined by a condition $(T-T_{c0})\ll T_{c0}$, where $T_{c0}$ is the critical temperature of a pristine system. In situations when it is suppressed, $T_c(x)<T_{c0}$, one has to consider generalizations as appropriate for the quantum regime of fluctuations $T<T_{c0}$. The existing theoretical approaches to this complicated problem of fluctuation-induced transport in disordered bulk superconductors are typically based on the pair-breaking scenario \cite{Ramazashvili:PRL97,Mineev:PRB01,Ikeda:JPSJ01,Lopatin:PRL05}.  In its simplest form $T_c$ suppression is described by the Abrikosov-Gor'kov equation \cite{AG:JETP61}
\begin{equation}\label{eq:AG}
\ln\frac{T}{T_{c0}}=\psi\left(\frac{1}{2}\right)-\psi\left(\frac{1}{2}+\frac{\Gamma}{2\pi T}\right), 
\end{equation}  
\noindent where $\psi(z)$ is the digamma function. At a given pair-breaking strength $\Gamma$, superconductivity is destroyed at $T=T_c(\Gamma)$, and conversely, at a given temperature $T$, at $\Gamma=\Gamma_c(T)$. In the BCS model, $\Gamma$ is related to the spin-flip scattering. In our estimates we will keep $\Gamma$ as the phenomenological parameter. Microscopically, it has to be determined from the proper model of SC-CDW interplay in the presence of disorder scattering, therefore, the actual form of Eq.\eqref{eq:AG} will be different in that case. In models where $\Gamma$ characterizes the rate of electron collisions with impurities, the estimation of Born scattering on the impurity potential $U$ gives for the scattering rate $\Gamma\sim\nu |U|^2n_{\text{imp}}$, where $\nu$ is the single particle density of  states and $n_{\text{imp}}$ is the impurity concentration. When irradiation acts as a source of point-like disorder, one may roughly assume $n_{\text{imp}}\propto x$. We notice from the data of Fig.\ref{fig:TcCDW-Dose} that $T_{c}(x)$ suppression in our samples is the most significant in the range of concentrations $x$ across the QCP. We therefore expect fluctuation to be especially pronounced in that part of the phase diagram. 

In applying the pair-breaking scenario to determine the fluctuation correction to the normal state conductivity, we follow the most recent diagrammatic calculation of Ref. \cite{AL:PRB23}, which can be straightforwardly generalized to the three-dimensional case. Focusing first on the temperature interval $T<T_{c0}$, it can be shown that the main contribution is determined by the Aslamazov-Larkin (AL) diagram, with the corresponding analytical expression for the conductivity correction given by 
\begin{equation}\label{eq:AL}
\delta\sigma=\left(\frac{e\nu D}{4\pi T}\right)^2\int\frac{q^2(\Re B'_{q,\omega})^2(\Im L^R_{q,\omega})^2}{T\sinh^2\frac{\omega}{2T}}d\Upsilon_{q,\omega}
\end{equation} 
\noindent where $D$ is the diffusion coefficient and integration spans the phase space $d\Upsilon_{q,\omega}=d\omega d^3q/(2\pi)^4$. In these notations, the vertex function, triangular block of the AL diagram, is given by 
\begin{equation}
B_{q,\omega}=\psi\left(\frac{1}{2}+\frac{\Gamma_q-i\omega}{4\pi T}\right),
\end{equation}
\noindent where $\Gamma_q=\Gamma+Dq^2$. The pair-propagator that describes proliferation of superconducting fluctuations, namely its retarded component, is given by \cite{Book}  
\begin{equation}
L^R_{q,\omega}=-\frac{\nu^{-1}}{\ln\frac{T}{T_{c0}}-\psi\left(\frac{1}{2}\right)+\psi\left(\frac{1}{2}+\frac{\Gamma_q-i\omega}{4\pi T}\right)}. 
\end{equation} 
The pole of $L_{q,\omega}$ at $\{Dq^2,\omega\}\to0$ traces the boundary
between the superconducting and normal phases. We analyze Eq.\eqref{eq:AL} analytically in two limits: (i) quantum fluctuations $T<\Xi<T_{c0}$, and (ii) thermal fluctuations $\Xi<T<T_{c0}$, where energy scale $\Xi=|\Gamma-\Gamma_c|$ measures the proximity to quantum critical regime.  

For case (i) one can neglect the frequency dependence of the vertex function, which implies $\Re B'_{q,\omega}\approx 4\pi T/\Gamma_q$. In the low-temperature limit, the propagator of fluctuations reduces to  
\begin{equation}
L^R_{q,\omega}\approx \nu^{-1}\ln^{-1}\left(\frac{\Gamma_c}{\Gamma_q-i\omega}\right), 
\end{equation} 
and it is sufficient to retain only the pole of the logarithm, since the branch-cut singularity yields a smaller contribution to the integral in Eq.\eqref{eq:AL}. This implies $\Im L^R_{q,\omega}\approx-\nu\Gamma_c\omega/(Dq^2+\Xi)^2$. These approximations for $B_{q,\omega}$ and $L^R_{q,\omega}$ taken together give a simplified form of Eq.\eqref{eq:AL}
\begin{equation}
\delta\sigma=\frac{(eD)^2}{T}\int\frac{q^2\omega^2\Gamma^2_cd\Upsilon_{q,\omega}}{\sinh^2\frac{\omega}{2T}(Dq^2+\Gamma)^2(Dq^2+\Xi)^4}.
\end{equation}  
Calculating the remaining integrals by noting that the most important domain of fluctuation momenta are $Dq^2\sim\Xi\ll\Gamma$, we obtain for $T<\Xi$ 
\begin{equation}\label{eq:AL-quantum}
\delta\sigma\simeq e^2\sqrt{\frac{\Gamma_c}{D}}\left(\frac{\Gamma_c}{\Xi}\right)^{\frac{3}{2}}\left(\frac{T}{\Gamma_c}\right)^2=\delta A_2 T^2.
\end{equation}
Here we omitted the numerical prefactor for brevity and retained only the main parametric dependence. 

In a similar manner, for case (ii) we can still neglect $\omega$ dependence of $B_{q,\omega}$, we must retain the full $\omega$ dependence of $L^R_{q,\omega}$, and we can also expand $\sinh(\omega/2T)\approx \omega/2T$, since in this limit $\{Dq^2,\omega\}\sim \Xi<T$. This leads us to the approximate expression instead of Eq.\eqref{eq:AL}
\begin{equation}
\delta\sigma=(eD)^2\int\frac{4Tq^2\Gamma^2_cd\Upsilon_{q,\omega}}{(Dq^2+\Gamma)^2((Dq^2+\Xi)^2+\omega^2)^2}.
\end{equation}  
After integration, one finds as a result for $\Xi<T<T_{c0}$
\begin{equation}\label{eq:AL-thermal}
\delta\sigma\simeq e^2\sqrt{\frac{\Gamma_c}{D}}\sqrt{\frac{\Gamma_c}{\Xi}}\frac{T}{\Gamma_c}=\delta A_1 T.
\end{equation} 
It is worth noting that this expression does not diverge at the critical point, since at low temperature $\Gamma_c\approx \Gamma_{c0}-\alpha T^2/\Gamma_{c0}$, where $\alpha\sim1$. 

These results should be contrasted to the behavior in the classical fluctuations regime $T\gtrsim T_{c0}$. In that case, one finds \cite{Book}
\begin{equation}
\delta\sigma=\left(\frac{\pi\nu eD}{8T}\right)^2\int\frac{q^2(\Im L^R_{q,\omega})^2}{3T\sinh^2\frac{\omega}{2T}}d\Upsilon_{q,\omega}
\end{equation} 
\noindent where
\begin{equation}
L^R_{q,\omega}\approx-\nu^{-1}\left[\frac{T-T_{c0}}{T}+\frac{\pi}{8T}(Dq^2-i\omega)\right]^{-1}.
\end{equation}
After integration, the correction can be reduced to the form 
\begin{equation}\label{eq:AL-classical}
\delta\sigma\simeq \frac{e^2}{\sqrt{D}}\frac{T}{\sqrt{T-T_{c0}}}, \quad T-T_{c0}\ll T_{c0} 
\end{equation}
As usual, the apparent divergence has to be regularized by the applicability of the perturbation theory, namely $\delta\sigma<\sigma$, where $
\sigma$ is the bare normal-state conductivity. This limits applicability of the above asymptotic behavior is to be outside of the critical region of strong fluctuations, $(T-T_{c0})/T_{c0}>Gi_3$. The Ginzburg number in 3D case can be estimated as $Gi_3\sim(\nu T_{c0}\xi^3)^{-2}\ll1$, where $\xi$ is the superconducting coherence length in the diffusive limit.  

Finally, at the much higher temperatures, $T\gg T_{c0}$, one may deduce from the available results \cite{Aslamazov:JLTP80,Altshuler:JETP83}
\begin{equation}
\delta\sigma\simeq - e^2\sqrt{\frac{T}{D}}\frac{1}{\ln(T/T_{c0})}, \quad T\gg T_{c0}. 
\end{equation}      
In this limit, however, fluctuations are dominated by the density of states and so-called regular part of the Maki-Thompson diagram. This is reflected in the opposite sign of the correction term, as compared to Eqs. \eqref{eq:AL-quantum}, \eqref{eq:AL-thermal}, and \eqref{eq:AL-classical}, thus increasing resistivity. 

This analysis suggests that superconducting fluctuations contribute to the $T-$linear and $T-$quadratic dependence of the resistivity observed in the experiment. Within the limitations of the perturbation theory, presented above, the corresponding corrections to resistivity can be presented as $\delta\rho=-\rho^2\delta\sigma$, where $\rho$ should be understood as a bare disorder-limited conductivity $\rho=A_0$. In this approximation, the coefficients $\delta A_{1,2}(\Gamma)$ from Eqs. \eqref{eq:AL-quantum} and \eqref{eq:AL-thermal} can be attributed to the renormalization of the fitting parameters $A_{1,2}$. The above model considerations indicate that once the system is tuned toward the quantum critical regime in concentration $x$, the regime of $T$-linear dependence expands, since $\Xi$ becomes smaller and the range of validity of Eq.\eqref{eq:AL-thermal} becomes wider. This is qualitatively consistent with the observed trend. Furthermore, the sign of the correction term from Eq.\eqref{eq:AL-thermal} is also consistent with the observed reduction in the $T$-linear slope of the resistivity governed by the coefficient $A_1$. The fact that the correction term changes sign underscores the point that the coefficients $A_{1,2}$ exhibit nonmonotonic behavior, seeing in Fig.\ref{fig:polynomial} . 

\section*{Conclusions}

In conclusion, it is shown that a controlled non-magnetic point-like disorder can tune the CDW/SC quasi-skutterudites (Ca$_{x}$Sr$_{1-x}$)$_{3}$(Rh, Ir)$_{4}$Sn$_{13}$ to and through the structural/CDW quantum critical point.
Starting at compositions below the known structural QCP at $x_c=0.9$, disorder induced by 2.5 MeV electron irradiation resulted in a progressively larger linear term and a reduced quadratic term in resistivity, $\rho\left(T\right)$, just above $T_c$. These novel experiments and qualitative theoretical analysis strongly support the general theoretical prediction that the non-magnetic disorder can drive the system toward the quantum critical regime.


\section*{Materials and Methods}

\textbf{Samples:} Single crystals (Ca$_{x}$Sr$_{1-x}$)$_{3}$Rh$_{4}$Sn$_{13}$ were
grown from tin flux \cite{Krieger2023}. A mixture of 3 parts (Ca + Sr of the desired ratio), 4 parts Rh and 93 parts Sn, was placed in an alumina crucible and sealed under vacuum in a quartz ampule along with a bit of quartz wool for filtration. The ampule was heated up to 1100~$^o$C, kept for six hours, fast cooled to 800~$^o$C and then cooled slowly to 490 $^o$C where the excess Sn was decanted in a centrifuge outside the furnace. The process is described elsewhere \cite{cairsn1}. The obtained samples were characterized by powder X-ray diffraction and energy-dispersive X-ray compositional analysis. This well-studied system shows good sample homogeneity and reproducibility of resistivity and EDX compositions between samples.

\textbf{Transport measurements:} Electrical resistivity was measured in a standard four-probe configuration in bar-shaped single crystals of typical length of 1-3 mm, width of 0.3-0.7 mm and thickness between 50 and 300 $\mu$m. To prepare the samples, the crystals were etched with HCl, cut to dimensions using a wire saw, and polished. The electrical contacts were formed by soldering 50 $\mu$m silver wires with tin-silver solder \cite{Krenkel2022,Krenkel2023} with typical contact resistances below 100 $\mu\Omega\cdot$cm.

\textbf{Electron irradiation:} The 2.5 MeV electron irradiation was performed at the ``SIRIUS'' accelerator facility at Laboratoire des Solides Irradi\'{e}s, \'{E}cole Polytechnique, Palaiseau, France. The acquired irradiation dose is measured by a calibrated Faraday trap behind the sample and is conveniently expressed in $\textrm{C}/\textrm{cm}^{2}$, where 1 $\textrm{C}/\textrm{cm}^{2}=6.24\times10^{18}\:\textrm{e}^{-}/\textrm{cm}^{2}$. Collisions of electrons with ions produce Frenkel pairs (vacancy + interstitial) because the relativistic energy transfer upon collisions matches well with the knock out energy threshold, typically between 20 and 80 eV. At these energies, the electron scattering cross-section is on the order of 100 barns. Other particles used for irradiation have either much lower cross-section (neutrons and gamma-rays), or may produce clusters (protons), and other correlated defects, including columnar tracks (heavy ions). 

Another important parameter for irradiation is the particles penetration depth into the material. While heavier particles require very thin samples because of beam attenuation, electrons produce relatively uniform disorder at mm-range depths for typical samples. Importantly, for the doses of irradiation studied, the produced defects are in the dilute limit, typically with about one defect per a thousands of host lattice ions. Furthermore, according to extensive Hall-effect measurements \cite{npjProzorov2019,ProzorovPhysRevX}, the electron irradiation does not ``charge-dope'' the samples and does not cause a shift of the chemical potential. 

The irradiation is conducted at low temperatures with the sample immersed in liquid hydrogen around 22 K. This is done to remove heat generated upon collisions and to prevent immediate recombination and clustering of the defects. Upon warming to room temperature, some pairs recombine, and some defects migrate to various sinks such as lattice defects, dislocations, and surfaces leaving a metastable population of point-like defects behind, typically about 70\% of the initially produced population. The final amount of the induced disorder is directly estimated from the changes in residual resistivity. Importantly, to avoid variation between different samples, in our experiments, the same sample was repeatedly irradiated and then measured to observe the progression of the results. Several compositions around the QCP concentration were studied. For a more detailed discussion of the use of controlled electron irradiation to study materials, the reader is referred to specialized books \cite{Damask1963,Thompson1969}.


\section*{Acknowledgments}

This work was supported by the National Science Foundation under Grant No. DMR-2219901. M.A.T. and K.R.J. were supported by the U.S. Department of Energy (DOE), Office of Science, Basic Energy Sciences, Materials Science and Engineering Division. Ames National Laboratory is operated for the U.S. DOE by Iowa State University under Contract No. DE-AC02-07CH11358. A.L. acknowledges the financial support by the National Science Foundation Grant No. DMR- 2203411 and a Research Fellowship funded by the Alexander von Humboldt Foundation. Work in France was supported by “Investissements d’Avenir” LabEx PALM (Grant No. ANR10-LABX-0039-PALM). The authors acknowledge support from the EMIR\&A French network (FR CNRS 3618) on the platform SIRIUS, proposals No. 20-5925 and 23-4663. Work at BNL (materials synthesis) was supported by the U.S. Department of Energy, Basic Energy Sciences, Division of Materials Science and Engineering, under Contract No. DE-SC0012704. C.P. acknowledges support from the Shanghai Key Laboratory of Material Frontiers Research in Extreme Environments, China (No. 22dz2260800) and Shanghai Science and Technology Committee, China (No. 22JC1410300).


\begin{thebibliography}{29}%
\makeatletter
\providecommand \@ifxundefined [1]{%
 \@ifx{#1\undefined}
}%
\providecommand \@ifnum [1]{%
 \ifnum #1\expandafter \@firstoftwo
 \else \expandafter \@secondoftwo
 \fi
}%
\providecommand \@ifx [1]{%
 \ifx #1\expandafter \@firstoftwo
 \else \expandafter \@secondoftwo
 \fi
}%
\providecommand \natexlab [1]{#1}%
\providecommand \enquote  [1]{``#1''}%
\providecommand \bibnamefont  [1]{#1}%
\providecommand \bibfnamefont [1]{#1}%
\providecommand \citenamefont [1]{#1}%
\providecommand \href@noop [0]{\@secondoftwo}%
\providecommand \href [0]{\begingroup \@sanitize@url \@href}%
\providecommand \@href[1]{\@@startlink{#1}\@@href}%
\providecommand \@@href[1]{\endgroup#1\@@endlink}%
\providecommand \@sanitize@url [0]{\catcode `\\12\catcode `\$12\catcode
  `\&12\catcode `\#12\catcode `\^12\catcode `\_12\catcode `\%12\relax}%
\providecommand \@@startlink[1]{}%
\providecommand \@@endlink[0]{}%
\providecommand \url  [0]{\begingroup\@sanitize@url \@url }%
\providecommand \@url [1]{\endgroup\@href {#1}{\urlprefix }}%
\providecommand \urlprefix  [0]{URL }%
\providecommand \Eprint [0]{\href }%
\providecommand \doibase [0]{https://doi.org/}%
\providecommand \selectlanguage [0]{\@gobble}%
\providecommand \bibinfo  [0]{\@secondoftwo}%
\providecommand \bibfield  [0]{\@secondoftwo}%
\providecommand \translation [1]{[#1]}%
\providecommand \BibitemOpen [0]{}%
\providecommand \bibitemStop [0]{}%
\providecommand \bibitemNoStop [0]{.\EOS\space}%
\providecommand \EOS [0]{\spacefactor3000\relax}%
\providecommand \BibitemShut  [1]{\csname bibitem#1\endcsname}%
\let\auto@bib@innerbib\@empty
\bibitem [{\citenamefont {Remeika}\ \emph {et~al.}(1980)\citenamefont
  {Remeika}, \citenamefont {Espinosa}, \citenamefont {Cooper}, \citenamefont
  {Barz}, \citenamefont {Rowell}, \citenamefont {McWhan}, \citenamefont
  {Vandenberg}, \citenamefont {Moncton}, \citenamefont {Fisk}, \citenamefont
  {Woolf}, \citenamefont {Hamaker}, \citenamefont {Maple}, \citenamefont
  {Shirane},\ and\ \citenamefont {Thomlinson}}]{Remeika1980}%
  \BibitemOpen
  \bibfield  {author} {\bibinfo {author} {\bibfnamefont {J.~P.}\ \bibnamefont
  {Remeika}}, \bibinfo {author} {\bibfnamefont {G.~P.}\ \bibnamefont
  {Espinosa}}, \bibinfo {author} {\bibfnamefont {A.~S.}\ \bibnamefont
  {Cooper}}, \bibinfo {author} {\bibfnamefont {H.}~\bibnamefont {Barz}},
  \bibinfo {author} {\bibfnamefont {J.~M.}\ \bibnamefont {Rowell}}, \bibinfo
  {author} {\bibfnamefont {D.~B.}\ \bibnamefont {McWhan}}, \bibinfo {author}
  {\bibfnamefont {J.~M.}\ \bibnamefont {Vandenberg}}, \bibinfo {author}
  {\bibfnamefont {D.~E.}\ \bibnamefont {Moncton}}, \bibinfo {author}
  {\bibfnamefont {Z.}~\bibnamefont {Fisk}}, \bibinfo {author} {\bibfnamefont
  {L.~D.}\ \bibnamefont {Woolf}}, \bibinfo {author} {\bibfnamefont {H.~C.}\
  \bibnamefont {Hamaker}}, \bibinfo {author} {\bibfnamefont {M.~B.}\
  \bibnamefont {Maple}}, \bibinfo {author} {\bibfnamefont {G.}~\bibnamefont
  {Shirane}},\ and\ \bibinfo {author} {\bibfnamefont {W.}~\bibnamefont
  {Thomlinson}},\ }\bibfield  {title} {\bibinfo {title} {A new family of
  ternary intermetallic superconducting/magnetic stannides},\ }\href
  {https://doi.org/10.1016/0038-1098(80)91099-6} {\bibfield  {journal}
  {\bibinfo  {journal} {Solid State Comm.}\ }\textbf {\bibinfo {volume} {34}},\
  \bibinfo {pages} {923} (\bibinfo {year} {1980})}\BibitemShut {NoStop}%
\bibitem [{\citenamefont {Klintberg}\ \emph {et~al.}(2012)\citenamefont
  {Klintberg}, \citenamefont {Goh}, \citenamefont {Alireza}, \citenamefont
  {Saines}, \citenamefont {Tompsett}, \citenamefont {Logg}, \citenamefont
  {Yang}, \citenamefont {Chen}, \citenamefont {Yoshimura},\ and\ \citenamefont
  {Grosche}}]{Klintberg2012}%
  \BibitemOpen
  \bibfield  {author} {\bibinfo {author} {\bibfnamefont {L.~E.}\ \bibnamefont
  {Klintberg}}, \bibinfo {author} {\bibfnamefont {S.~K.}\ \bibnamefont {Goh}},
  \bibinfo {author} {\bibfnamefont {P.~L.}\ \bibnamefont {Alireza}}, \bibinfo
  {author} {\bibfnamefont {P.~J.}\ \bibnamefont {Saines}}, \bibinfo {author}
  {\bibfnamefont {D.~A.}\ \bibnamefont {Tompsett}}, \bibinfo {author}
  {\bibfnamefont {P.~W.}\ \bibnamefont {Logg}}, \bibinfo {author}
  {\bibfnamefont {J.}~\bibnamefont {Yang}}, \bibinfo {author} {\bibfnamefont
  {B.}~\bibnamefont {Chen}}, \bibinfo {author} {\bibfnamefont {K.}~\bibnamefont
  {Yoshimura}},\ and\ \bibinfo {author} {\bibfnamefont {F.~M.}\ \bibnamefont
  {Grosche}},\ }\bibfield  {title} {\bibinfo {title} {Pressure and composition
  induced structural quantum phase transition in the cubic superconductor
  {(Sr,Ca)$_3$Ir$_4$Sn$_{13}$}},\ }\href@noop {} {\bibfield  {journal}
  {\bibinfo  {journal} {Phys. Rev. Lett.}\ }\textbf {\bibinfo {volume} {109}},\
  \bibinfo {pages} {237008} (\bibinfo {year} {2012})}\BibitemShut {NoStop}%
\bibitem [{\citenamefont {Goh}\ \emph {et~al.}(2015)\citenamefont {Goh},
  \citenamefont {Tompsett}, \citenamefont {Saines}, \citenamefont {Chang},
  \citenamefont {Matsumoto}, \citenamefont {Imai}, \citenamefont {Yoshimura},\
  and\ \citenamefont {Grosche}}]{Goh2015}%
  \BibitemOpen
  \bibfield  {author} {\bibinfo {author} {\bibfnamefont {S.~K.}\ \bibnamefont
  {Goh}}, \bibinfo {author} {\bibfnamefont {D.~A.}\ \bibnamefont {Tompsett}},
  \bibinfo {author} {\bibfnamefont {P.~J.}\ \bibnamefont {Saines}}, \bibinfo
  {author} {\bibfnamefont {H.~C.}\ \bibnamefont {Chang}}, \bibinfo {author}
  {\bibfnamefont {T.}~\bibnamefont {Matsumoto}}, \bibinfo {author}
  {\bibfnamefont {M.}~\bibnamefont {Imai}}, \bibinfo {author} {\bibfnamefont
  {K.}~\bibnamefont {Yoshimura}},\ and\ \bibinfo {author} {\bibfnamefont
  {F.~M.}\ \bibnamefont {Grosche}},\ }\bibfield  {title} {\bibinfo {title}
  {Ambient pressure structural quantum critical point in the phase diagram of
  {(Ca$_x$Sr$_{1-x}$)$_3$Rh$_4$Sn$_{13}$}},\ }\href
  {https://doi.org/10.1103/PhysRevLett.114.097002} {\bibfield  {journal}
  {\bibinfo  {journal} {Phys. Rev. Lett.}\ }\textbf {\bibinfo {volume} {114}},\
  \bibinfo {pages} {097002} (\bibinfo {year} {2015})}\BibitemShut {NoStop}%
\bibitem [{\citenamefont {Terasaki}\ \emph {et~al.}(2021)\citenamefont
  {Terasaki}, \citenamefont {Yamaguchi}, \citenamefont {Ishii}, \citenamefont
  {Tada}, \citenamefont {Yamamoto},\ and\ \citenamefont {Mori}}]{Teraski2021}%
  \BibitemOpen
  \bibfield  {author} {\bibinfo {author} {\bibfnamefont {Y.}~\bibnamefont
  {Terasaki}}, \bibinfo {author} {\bibfnamefont {R.}~\bibnamefont {Yamaguchi}},
  \bibinfo {author} {\bibfnamefont {Y.}~\bibnamefont {Ishii}}, \bibinfo
  {author} {\bibfnamefont {Y.}~\bibnamefont {Tada}}, \bibinfo {author}
  {\bibfnamefont {A.}~\bibnamefont {Yamamoto}},\ and\ \bibinfo {author}
  {\bibfnamefont {S.}~\bibnamefont {Mori}},\ }\bibfield  {title} {\bibinfo
  {title} {Superconductivity enhanced by abundant low-energy phonons in
  {(Sr$_{1-x}$Ca$_x$)$_3$Rh$_4$Sn$_{13}$}},\ }\href
  {https://doi.org/10.7566/JPSJ.90.113704} {\bibfield  {journal} {\bibinfo
  {journal} {Journal of the Physical Society of Japan}\ }\textbf {\bibinfo
  {volume} {90}},\ \bibinfo {pages} {113704} (\bibinfo {year}
  {2021})}\BibitemShut {NoStop}%
\bibitem [{\citenamefont {Veiga}\ \emph {et~al.}(2020)\citenamefont {Veiga},
  \citenamefont {Mardegan}, \citenamefont {v.~Zimmermann}, \citenamefont
  {Maimone}, \citenamefont {Carneiro}, \citenamefont {Fontes}, \citenamefont
  {Strempfer}, \citenamefont {Granado}, \citenamefont {Pagliuso},\ and\
  \citenamefont {Bittar}}]{Veiga2020}%
  \BibitemOpen
  \bibfield  {author} {\bibinfo {author} {\bibfnamefont {L.~S.~I.}\
  \bibnamefont {Veiga}}, \bibinfo {author} {\bibfnamefont {J.~R.~L.}\
  \bibnamefont {Mardegan}}, \bibinfo {author} {\bibfnamefont {M.}~\bibnamefont
  {v.~Zimmermann}}, \bibinfo {author} {\bibfnamefont {D.~T.}\ \bibnamefont
  {Maimone}}, \bibinfo {author} {\bibfnamefont {F.~B.}\ \bibnamefont
  {Carneiro}}, \bibinfo {author} {\bibfnamefont {M.~B.}\ \bibnamefont
  {Fontes}}, \bibinfo {author} {\bibfnamefont {J.}~\bibnamefont {Strempfer}},
  \bibinfo {author} {\bibfnamefont {E.}~\bibnamefont {Granado}}, \bibinfo
  {author} {\bibfnamefont {P.~G.}\ \bibnamefont {Pagliuso}},\ and\ \bibinfo
  {author} {\bibfnamefont {E.~M.}\ \bibnamefont {Bittar}},\ }\bibfield  {title}
  {\bibinfo {title} {Possible quantum fluctuations in the vicinity of the
  quantum critical point of {(Sr,Ca)$_3$Ir$_4$Sn$_{13}$} revealed by
  high-energy x-ray diffraction},\ }\href
  {https://doi.org/10.1103/PhysRevB.101.104511} {\bibfield  {journal} {\bibinfo
   {journal} {Phys. Rev. B}\ }\textbf {\bibinfo {volume} {101}},\ \bibinfo
  {pages} {104511} (\bibinfo {year} {2020})}\BibitemShut {NoStop}%
\bibitem [{\citenamefont {Carneiro}\ \emph {et~al.}(2020)\citenamefont
  {Carneiro}, \citenamefont {Veiga}, \citenamefont {Mardegan}, \citenamefont
  {Khan}, \citenamefont {Macchiutti}, \citenamefont {L\'opez},\ and\
  \citenamefont {Bittar}}]{Carneiro2020}%
  \BibitemOpen
  \bibfield  {author} {\bibinfo {author} {\bibfnamefont {F.~B.}\ \bibnamefont
  {Carneiro}}, \bibinfo {author} {\bibfnamefont {L.~S.~I.}\ \bibnamefont
  {Veiga}}, \bibinfo {author} {\bibfnamefont {J.~R.~L.}\ \bibnamefont
  {Mardegan}}, \bibinfo {author} {\bibfnamefont {R.}~\bibnamefont {Khan}},
  \bibinfo {author} {\bibfnamefont {C.}~\bibnamefont {Macchiutti}}, \bibinfo
  {author} {\bibfnamefont {A.}~\bibnamefont {L\'opez}},\ and\ \bibinfo {author}
  {\bibfnamefont {E.~M.}\ \bibnamefont {Bittar}},\ }\bibfield  {title}
  {\bibinfo {title} {Unveiling charge density wave quantum phase transitions by
  x-ray diffraction},\ }\href {https://doi.org/10.1103/PhysRevB.101.195135}
  {\bibfield  {journal} {\bibinfo  {journal} {Phys. Rev. B}\ }\textbf {\bibinfo
  {volume} {101}},\ \bibinfo {pages} {195135} (\bibinfo {year}
  {2020})}\BibitemShut {NoStop}%
\bibitem [{\citenamefont {Krieger}\ \emph {et~al.}(2023)\citenamefont
  {Krieger}, \citenamefont {Guguchia}, \citenamefont {Khasanov}, \citenamefont
  {Biswas}, \citenamefont {Li}, \citenamefont {Wang}, \citenamefont
  {Petrovic},\ and\ \citenamefont {Morenzoni}}]{Krieger2023}%
  \BibitemOpen
  \bibfield  {author} {\bibinfo {author} {\bibfnamefont {J.~A.}\ \bibnamefont
  {Krieger}}, \bibinfo {author} {\bibfnamefont {Z.}~\bibnamefont {Guguchia}},
  \bibinfo {author} {\bibfnamefont {R.}~\bibnamefont {Khasanov}}, \bibinfo
  {author} {\bibfnamefont {P.~K.}\ \bibnamefont {Biswas}}, \bibinfo {author}
  {\bibfnamefont {L.}~\bibnamefont {Li}}, \bibinfo {author} {\bibfnamefont
  {K.}~\bibnamefont {Wang}}, \bibinfo {author} {\bibfnamefont {C.}~\bibnamefont
  {Petrovic}},\ and\ \bibinfo {author} {\bibfnamefont {E.}~\bibnamefont
  {Morenzoni}},\ }\bibfield  {title} {\bibinfo {title} {Enhancement of
  superconductivity at a quantum critical point in
  {(Ca$_x$Sr$_{1-x}$)$_3$Rh$_4$Sn$_{13}$}},\ }\href
  {https://doi.org/10.1088/1742-6596/2462/1/012060} {\bibfield  {journal}
  {\bibinfo  {journal} {Journal of Physics: Conference Series}\ }\textbf
  {\bibinfo {volume} {2462}},\ \bibinfo {pages} {012060} (\bibinfo {year}
  {2023})}\BibitemShut {NoStop}%
\bibitem [{\citenamefont {Liu}\ \emph {et~al.}(2022)\citenamefont {Liu},
  \citenamefont {Zhang}, \citenamefont {Lai}, \citenamefont {Moriyama},
  \citenamefont {Tallon}, \citenamefont {Yoshimura},\ and\ \citenamefont
  {Goh}}]{Liu2022}%
  \BibitemOpen
  \bibfield  {author} {\bibinfo {author} {\bibfnamefont {X.}~\bibnamefont
  {Liu}}, \bibinfo {author} {\bibfnamefont {W.}~\bibnamefont {Zhang}}, \bibinfo
  {author} {\bibfnamefont {K.~T.}\ \bibnamefont {Lai}}, \bibinfo {author}
  {\bibfnamefont {K.}~\bibnamefont {Moriyama}}, \bibinfo {author}
  {\bibfnamefont {J.~L.}\ \bibnamefont {Tallon}}, \bibinfo {author}
  {\bibfnamefont {K.}~\bibnamefont {Yoshimura}},\ and\ \bibinfo {author}
  {\bibfnamefont {S.~K.}\ \bibnamefont {Goh}},\ }\bibfield  {title} {\bibinfo
  {title} {Peak in the critical current density in
  {(Ca$_x$Sr$_{1-x}$)$_3$Rh$_4$Sn$_{13}$} tuned towards the structural quantum
  critical point},\ }\href {https://doi.org/10.1103/PhysRevB.105.214524}
  {\bibfield  {journal} {\bibinfo  {journal} {Phys. Rev. B}\ }\textbf {\bibinfo
  {volume} {105}},\ \bibinfo {pages} {214524} (\bibinfo {year}
  {2022})}\BibitemShut {NoStop}%
\bibitem [{\citenamefont {Krenkel}\ \emph {et~al.}(2024)\citenamefont
  {Krenkel}, \citenamefont {Tanatar}, \citenamefont {Ghimire}, \citenamefont
  {Joshi}, \citenamefont {Chen}, \citenamefont {Petrovic},\ and\ \citenamefont
  {Prozorov}}]{Krenkel2023}%
  \BibitemOpen
  \bibfield  {author} {\bibinfo {author} {\bibfnamefont {E.~H.}\ \bibnamefont
  {Krenkel}}, \bibinfo {author} {\bibfnamefont {M.~A.}\ \bibnamefont
  {Tanatar}}, \bibinfo {author} {\bibfnamefont {S.}~\bibnamefont {Ghimire}},
  \bibinfo {author} {\bibfnamefont {K.~R.}\ \bibnamefont {Joshi}}, \bibinfo
  {author} {\bibfnamefont {S.}~\bibnamefont {Chen}}, \bibinfo {author}
  {\bibfnamefont {C.}~\bibnamefont {Petrovic}},\ and\ \bibinfo {author}
  {\bibfnamefont {R.}~\bibnamefont {Prozorov}},\ }\bibfield  {title} {\bibinfo
  {title} {{Robust superconductivity and the suppression of charge-density wave
  in the quasi-skutterudites
  $\text{Ca}_{3}(\text{Ir}_{1-x}\text{Rh}_{x})_{4}\text{Sn}_{13}$ single
  crystals at ambient pressure}},\ }\href
  {https://doi.org/10.1088/1361-648x/ad5485} {\bibfield  {journal} {\bibinfo
  {journal} {J. Phys. Condens. Matter}\ }\textbf {\bibinfo {volume} {1}},\
  \bibinfo {pages} {1} (\bibinfo {year} {2024})}\BibitemShut {NoStop}%
\bibitem [{\citenamefont {Belitz}\ and\ \citenamefont
  {Vojta}(2005)}]{Belitz2005}%
  \BibitemOpen
  \bibfield  {author} {\bibinfo {author} {\bibfnamefont {D.}~\bibnamefont
  {Belitz}}\ and\ \bibinfo {author} {\bibfnamefont {T.}~\bibnamefont {Vojta}},\
  }\bibfield  {title} {\bibinfo {title} {{How generic scale invariance
  influences quantum and classical phase transitions}},\ }\href
  {https://doi.org/10.1103/RevModPhys.77.579} {\bibfield  {journal} {\bibinfo
  {journal} {Rev. Mod. Phys.}\ }\textbf {\bibinfo {volume} {77}},\ \bibinfo
  {pages} {579} (\bibinfo {year} {2005})}\BibitemShut {NoStop}%
\bibitem [{\citenamefont {Joshi}\ \emph {et~al.}(2020)\citenamefont {Joshi},
  \citenamefont {Nusran}, \citenamefont {Tanatar}, \citenamefont {Cho},
  \citenamefont {Bud'ko}, \citenamefont {Canfield}, \citenamefont {Fernandes},
  \citenamefont {Levchenko},\ and\ \citenamefont {Prozorov}}]{Joshi2020}%
  \BibitemOpen
  \bibfield  {author} {\bibinfo {author} {\bibfnamefont {K.~R.}\ \bibnamefont
  {Joshi}}, \bibinfo {author} {\bibfnamefont {N.~M.}\ \bibnamefont {Nusran}},
  \bibinfo {author} {\bibfnamefont {M.~A.}\ \bibnamefont {Tanatar}}, \bibinfo
  {author} {\bibfnamefont {K.}~\bibnamefont {Cho}}, \bibinfo {author}
  {\bibfnamefont {S.~L.}\ \bibnamefont {Bud'ko}}, \bibinfo {author}
  {\bibfnamefont {P.~C.}\ \bibnamefont {Canfield}}, \bibinfo {author}
  {\bibfnamefont {R.~M.}\ \bibnamefont {Fernandes}}, \bibinfo {author}
  {\bibfnamefont {A.}~\bibnamefont {Levchenko}},\ and\ \bibinfo {author}
  {\bibfnamefont {R.}~\bibnamefont {Prozorov}},\ }\bibfield  {title} {\bibinfo
  {title} {{Quantum phase transition inside the superconducting dome of
  Ba(Fe$_{1-x}$Co$_{x}$)$_{2}$As$_{2}$ from diamond-based optical
  magnetometry}},\ }\href {https://doi.org/10.1088/1367-2630/ab85a9} {\bibfield
   {journal} {\bibinfo  {journal} {New J. Phys.}\ }\textbf {\bibinfo {volume}
  {22}},\ \bibinfo {pages} {053037} (\bibinfo {year} {2020})}\BibitemShut
  {NoStop}%
\bibitem [{\citenamefont {Hashimoto}\ \emph {et~al.}(2012)\citenamefont
  {Hashimoto}, \citenamefont {Cho}, \citenamefont {Shibauchi}, \citenamefont
  {Kasahara}, \citenamefont {Mizukami}, \citenamefont {Katsumata},
  \citenamefont {Tsuruhara}, \citenamefont {Terashima}, \citenamefont {Ikeda},
  \citenamefont {Tanatar}, \citenamefont {Kitano}, \citenamefont {Salovich},
  \citenamefont {Giannetta}, \citenamefont {Walmsley}, \citenamefont
  {Carrington}, \citenamefont {Prozorov},\ and\ \citenamefont
  {Matsuda}}]{Hashimoto2012}%
  \BibitemOpen
  \bibfield  {author} {\bibinfo {author} {\bibfnamefont {K.}~\bibnamefont
  {Hashimoto}}, \bibinfo {author} {\bibfnamefont {K.}~\bibnamefont {Cho}},
  \bibinfo {author} {\bibfnamefont {T.}~\bibnamefont {Shibauchi}}, \bibinfo
  {author} {\bibfnamefont {S.}~\bibnamefont {Kasahara}}, \bibinfo {author}
  {\bibfnamefont {Y.}~\bibnamefont {Mizukami}}, \bibinfo {author}
  {\bibfnamefont {R.}~\bibnamefont {Katsumata}}, \bibinfo {author}
  {\bibfnamefont {Y.}~\bibnamefont {Tsuruhara}}, \bibinfo {author}
  {\bibfnamefont {T.}~\bibnamefont {Terashima}}, \bibinfo {author}
  {\bibfnamefont {H.}~\bibnamefont {Ikeda}}, \bibinfo {author} {\bibfnamefont
  {M.~A.}\ \bibnamefont {Tanatar}}, \bibinfo {author} {\bibfnamefont
  {H.}~\bibnamefont {Kitano}}, \bibinfo {author} {\bibfnamefont
  {N.}~\bibnamefont {Salovich}}, \bibinfo {author} {\bibfnamefont {R.~W.}\
  \bibnamefont {Giannetta}}, \bibinfo {author} {\bibfnamefont {P.}~\bibnamefont
  {Walmsley}}, \bibinfo {author} {\bibfnamefont {A.}~\bibnamefont
  {Carrington}}, \bibinfo {author} {\bibfnamefont {R.}~\bibnamefont
  {Prozorov}},\ and\ \bibinfo {author} {\bibfnamefont {Y.}~\bibnamefont
  {Matsuda}},\ }\bibfield  {title} {\bibinfo {title} {{A Sharp Peak of the
  Zero-Temperature Penetration Depth at Optimal Composition in
  BaFe$_2$(As$_{1-x}$P$_x$)$_2$}},\ }\href
  {https://doi.org/10.1126/science.1219821} {\bibfield  {journal} {\bibinfo
  {journal} {Science}\ }\textbf {\bibinfo {volume} {336}},\ \bibinfo {pages}
  {1554} (\bibinfo {year} {2012})}\BibitemShut {NoStop}%
\bibitem [{\citenamefont {Krenkel}\ \emph {et~al.}(2022)\citenamefont
  {Krenkel}, \citenamefont {Tanatar}, \citenamefont {Ko\'{n}czykowski},
  \citenamefont {Grasset}, \citenamefont {Timmons}, \citenamefont {Ghimire},
  \citenamefont {Joshi}, \citenamefont {Lee}, \citenamefont {Ke}, \citenamefont
  {Chen}, \citenamefont {Petrovic}, \citenamefont {Orth}, \citenamefont
  {Scheurer},\ and\ \citenamefont {Prozorov}}]{Krenkel2022}%
  \BibitemOpen
  \bibfield  {author} {\bibinfo {author} {\bibfnamefont {E.~H.}\ \bibnamefont
  {Krenkel}}, \bibinfo {author} {\bibfnamefont {M.~A.}\ \bibnamefont
  {Tanatar}}, \bibinfo {author} {\bibfnamefont {M.}~\bibnamefont
  {Ko\'{n}czykowski}}, \bibinfo {author} {\bibfnamefont {R.}~\bibnamefont
  {Grasset}}, \bibinfo {author} {\bibfnamefont {E.~I.}\ \bibnamefont
  {Timmons}}, \bibinfo {author} {\bibfnamefont {S.}~\bibnamefont {Ghimire}},
  \bibinfo {author} {\bibfnamefont {K.~R.}\ \bibnamefont {Joshi}}, \bibinfo
  {author} {\bibfnamefont {Y.}~\bibnamefont {Lee}}, \bibinfo {author}
  {\bibfnamefont {L.}~\bibnamefont {Ke}}, \bibinfo {author} {\bibfnamefont
  {S.}~\bibnamefont {Chen}}, \bibinfo {author} {\bibfnamefont {C.}~\bibnamefont
  {Petrovic}}, \bibinfo {author} {\bibfnamefont {P.~P.}\ \bibnamefont {Orth}},
  \bibinfo {author} {\bibfnamefont {M.~S.}\ \bibnamefont {Scheurer}},\ and\
  \bibinfo {author} {\bibfnamefont {R.}~\bibnamefont {Prozorov}},\ }\bibfield
  {title} {\bibinfo {title} {Possible unconventional pairing in
  {(Ca,Sr)}$_{3}${(Rh,Ir)}$_{4}${Sn}$_{13}$ superconductors revealed by
  controlling disorder},\ }\href {https://doi.org/10.1103/PhysRevB.105.094521}
  {\bibfield  {journal} {\bibinfo  {journal} {Phys. Rev. B}\ }\textbf {\bibinfo
  {volume} {105}},\ \bibinfo {pages} {094521} (\bibinfo {year}
  {2022})}\BibitemShut {NoStop}%
\bibitem [{\citenamefont {Cho}\ \emph {et~al.}(2018)\citenamefont {Cho},
  \citenamefont {Ko{\'{n}}czykowski}, \citenamefont {Teknowijoyo},
  \citenamefont {Tanatar}, \citenamefont {Guss}, \citenamefont {Gartin},
  \citenamefont {Wilde}, \citenamefont {Kreyssig}, \citenamefont {McQueeney},
  \citenamefont {Goldman}, \citenamefont {Mishra}, \citenamefont {Hirschfeld},\
  and\ \citenamefont {Prozorov}}]{KyuilNbSe2}%
  \BibitemOpen
  \bibfield  {author} {\bibinfo {author} {\bibfnamefont {K.}~\bibnamefont
  {Cho}}, \bibinfo {author} {\bibfnamefont {M.}~\bibnamefont
  {Ko{\'{n}}czykowski}}, \bibinfo {author} {\bibfnamefont {S.}~\bibnamefont
  {Teknowijoyo}}, \bibinfo {author} {\bibfnamefont {M.~A.}\ \bibnamefont
  {Tanatar}}, \bibinfo {author} {\bibfnamefont {J.}~\bibnamefont {Guss}},
  \bibinfo {author} {\bibfnamefont {P.~B.}\ \bibnamefont {Gartin}}, \bibinfo
  {author} {\bibfnamefont {J.~M.}\ \bibnamefont {Wilde}}, \bibinfo {author}
  {\bibfnamefont {A.}~\bibnamefont {Kreyssig}}, \bibinfo {author}
  {\bibfnamefont {R.~J.}\ \bibnamefont {McQueeney}}, \bibinfo {author}
  {\bibfnamefont {A.~I.}\ \bibnamefont {Goldman}}, \bibinfo {author}
  {\bibfnamefont {V.}~\bibnamefont {Mishra}}, \bibinfo {author} {\bibfnamefont
  {P.~J.}\ \bibnamefont {Hirschfeld}},\ and\ \bibinfo {author} {\bibfnamefont
  {R.}~\bibnamefont {Prozorov}},\ }\bibfield  {title} {\bibinfo {title} {Using
  controlled disorder to probe the interplay between charge order and
  superconductivity in {NbSe$_2$}},\ }\href
  {https://doi.org/10.1038/s41467-018-05153-0} {\bibfield  {journal} {\bibinfo
  {journal} {Nat. Comm.}\ }\textbf {\bibinfo {volume} {9}},\ \bibinfo {pages}
  {2796} (\bibinfo {year} {2018})}\BibitemShut {NoStop}%
\bibitem [{\citenamefont {Machida}(1984)}]{Machida:JPSJ84}%
  \BibitemOpen
  \bibfield  {author} {\bibinfo {author} {\bibfnamefont {K.}~\bibnamefont
  {Machida}},\ }\bibfield  {title} {\bibinfo {title} {Charge density wave and
  superconductivity in anisotropic materials},\ }\href
  {https://doi.org/10.1143/JPSJ.53.712} {\bibfield  {journal} {\bibinfo
  {journal} {J. Phys. Soc. Japan}\ }\textbf {\bibinfo {volume} {53}},\ \bibinfo
  {pages} {712} (\bibinfo {year} {1984})}\BibitemShut {NoStop}%
\bibitem [{\citenamefont {Larkin}\ and\ \citenamefont {Varlamov}(2005)}]{Book}%
  \BibitemOpen
  \bibfield  {author} {\bibinfo {author} {\bibfnamefont {A.}~\bibnamefont
  {Larkin}}\ and\ \bibinfo {author} {\bibfnamefont {A.}~\bibnamefont
  {Varlamov}},\ }\href
  {https://doi.org/10.1093/acprof:oso/9780198528159.001.0001} {\emph {\bibinfo
  {title} {{Theory of Fluctuations in Superconductors}}}}\ (\bibinfo
  {publisher} {Oxford University Press},\ \bibinfo {year} {2005})\BibitemShut
  {NoStop}%
\bibitem [{\citenamefont {Ramazashvili}\ and\ \citenamefont
  {Coleman}(1997)}]{Ramazashvili:PRL97}%
  \BibitemOpen
  \bibfield  {author} {\bibinfo {author} {\bibfnamefont {R.}~\bibnamefont
  {Ramazashvili}}\ and\ \bibinfo {author} {\bibfnamefont {P.}~\bibnamefont
  {Coleman}},\ }\bibfield  {title} {\bibinfo {title} {{Superconducting Quantum
  Critical Point}},\ }\href {https://doi.org/10.1103/PhysRevLett.79.3752}
  {\bibfield  {journal} {\bibinfo  {journal} {Phys. Rev. Lett.}\ }\textbf
  {\bibinfo {volume} {79}},\ \bibinfo {pages} {3752} (\bibinfo {year}
  {1997})}\BibitemShut {NoStop}%
\bibitem [{\citenamefont {Mineev}\ and\ \citenamefont
  {Sigrist}(2001)}]{Mineev:PRB01}%
  \BibitemOpen
  \bibfield  {author} {\bibinfo {author} {\bibfnamefont {V.~P.}\ \bibnamefont
  {Mineev}}\ and\ \bibinfo {author} {\bibfnamefont {M.}~\bibnamefont
  {Sigrist}},\ }\bibfield  {title} {\bibinfo {title} {Critical fluctuation
  effects near the normal-metal--superconductor phase transition at low
  temperatures},\ }\href {https://doi.org/10.1103/PhysRevB.63.172504}
  {\bibfield  {journal} {\bibinfo  {journal} {Phys. Rev. B}\ }\textbf {\bibinfo
  {volume} {63}},\ \bibinfo {pages} {172504} (\bibinfo {year}
  {2001})}\BibitemShut {NoStop}%
\bibitem [{\citenamefont {Adachi}\ and\ \citenamefont
  {Ikeda}(2001)}]{Ikeda:JPSJ01}%
  \BibitemOpen
  \bibfield  {author} {\bibinfo {author} {\bibfnamefont {H.}~\bibnamefont
  {Adachi}}\ and\ \bibinfo {author} {\bibfnamefont {R.}~\bibnamefont {Ikeda}},\
  }\bibfield  {title} {\bibinfo {title} {Fluctuation conductivity in
  unconventional superconductors near critical disorder},\ }\href
  {https://doi.org/10.1143/JPSJ.70.2848} {\bibfield  {journal} {\bibinfo
  {journal} {Journal of the Physical Society of Japan}\ }\textbf {\bibinfo
  {volume} {70}},\ \bibinfo {pages} {2848} (\bibinfo {year}
  {2001})}\BibitemShut {NoStop}%
\bibitem [{\citenamefont {Lopatin}\ \emph {et~al.}(2005)\citenamefont
  {Lopatin}, \citenamefont {Shah},\ and\ \citenamefont
  {Vinokur}}]{Lopatin:PRL05}%
  \BibitemOpen
  \bibfield  {author} {\bibinfo {author} {\bibfnamefont {A.~V.}\ \bibnamefont
  {Lopatin}}, \bibinfo {author} {\bibfnamefont {N.}~\bibnamefont {Shah}},\ and\
  \bibinfo {author} {\bibfnamefont {V.~M.}\ \bibnamefont {Vinokur}},\
  }\bibfield  {title} {\bibinfo {title} {Fluctuation conductivity of thin films
  and nanowires near a parallel-field-tuned superconducting quantum phase
  transition},\ }\href {https://doi.org/10.1103/PhysRevLett.94.037003}
  {\bibfield  {journal} {\bibinfo  {journal} {Phys. Rev. Lett.}\ }\textbf
  {\bibinfo {volume} {94}},\ \bibinfo {pages} {037003} (\bibinfo {year}
  {2005})}\BibitemShut {NoStop}%
\bibitem [{\citenamefont {Abrikosov}\ and\ \citenamefont
  {Gor'kov}(1961)}]{AG:JETP61}%
  \BibitemOpen
  \bibfield  {author} {\bibinfo {author} {\bibfnamefont {A.~A.}\ \bibnamefont
  {Abrikosov}}\ and\ \bibinfo {author} {\bibfnamefont {L.~P.}\ \bibnamefont
  {Gor'kov}},\ }\bibfield  {title} {\bibinfo {title} {Contribution to the
  theory of superconducting alloys with paramagnetic impurities},\ }\href@noop
  {} {\bibfield  {journal} {\bibinfo  {journal} {Sov. Phys. JETP}\ }\textbf
  {\bibinfo {volume} {12}},\ \bibinfo {pages} {1243} (\bibinfo {year}
  {1961})}\BibitemShut {NoStop}%
\bibitem [{\citenamefont {Dzero}\ \emph {et~al.}(2023)\citenamefont {Dzero},
  \citenamefont {Khodas},\ and\ \citenamefont {Levchenko}}]{AL:PRB23}%
  \BibitemOpen
  \bibfield  {author} {\bibinfo {author} {\bibfnamefont {M.}~\bibnamefont
  {Dzero}}, \bibinfo {author} {\bibfnamefont {M.}~\bibnamefont {Khodas}},\ and\
  \bibinfo {author} {\bibfnamefont {A.}~\bibnamefont {Levchenko}},\ }\bibfield
  {title} {\bibinfo {title} {Transport anomalies in multiband superconductors
  near the quantum critical point},\ }\href
  {https://doi.org/10.1103/PhysRevB.108.184513} {\bibfield  {journal} {\bibinfo
   {journal} {Phys. Rev. B}\ }\textbf {\bibinfo {volume} {108}},\ \bibinfo
  {pages} {184513} (\bibinfo {year} {2023})}\BibitemShut {NoStop}%
\bibitem [{\citenamefont {Aslamazov}\ and\ \citenamefont
  {Varlamov}(1980)}]{Aslamazov:JLTP80}%
  \BibitemOpen
  \bibfield  {author} {\bibinfo {author} {\bibfnamefont {L.~G.}\ \bibnamefont
  {Aslamazov}}\ and\ \bibinfo {author} {\bibfnamefont {A.~A.}\ \bibnamefont
  {Varlamov}},\ }\bibfield  {title} {\bibinfo {title} {Fluctuation conductivity
  in intercalated superconductors},\ }\href
  {https://doi.org/10.1007/BF00115277} {\bibfield  {journal} {\bibinfo
  {journal} {Journal of Low Temperature Physics}\ }\textbf {\bibinfo {volume}
  {38}},\ \bibinfo {pages} {223} (\bibinfo {year} {1980})}\BibitemShut
  {NoStop}%
\bibitem [{\citenamefont {Al'tshuler}\ \emph {et~al.}(1983)\citenamefont
  {Al'tshuler}, \citenamefont {Varlamov},\ and\ \citenamefont
  {Reizer}}]{Altshuler:JETP83}%
  \BibitemOpen
  \bibfield  {author} {\bibinfo {author} {\bibfnamefont {B.~L.}\ \bibnamefont
  {Al'tshuler}}, \bibinfo {author} {\bibfnamefont {A.~A.}\ \bibnamefont
  {Varlamov}},\ and\ \bibinfo {author} {\bibfnamefont {M.~Y.}\ \bibnamefont
  {Reizer}},\ }\bibfield  {title} {\bibinfo {title} {Interelectron effects and
  the conductivity of disordered twodimensional electron systems},\ }\href@noop
  {} {\bibfield  {journal} {\bibinfo  {journal} {Sov. Phys. JETP}\ }\textbf
  {\bibinfo {volume} {57}},\ \bibinfo {pages} {1329} (\bibinfo {year}
  {1983})}\BibitemShut {NoStop}%
\bibitem [{\citenamefont {Wang}\ and\ \citenamefont
  {Petrovic}(2012)}]{cairsn1}%
  \BibitemOpen
  \bibfield  {author} {\bibinfo {author} {\bibfnamefont {K.}~\bibnamefont
  {Wang}}\ and\ \bibinfo {author} {\bibfnamefont {C.}~\bibnamefont
  {Petrovic}},\ }\bibfield  {title} {\bibinfo {title}
  {{Ca$_{3}$Ir$_{4}$Sn$_{13}$}: A weakly correlated nodeless superconductor},\
  }\href {https://doi.org/10.1103/PhysRevB.86.024522} {\bibfield  {journal}
  {\bibinfo  {journal} {Phys. Rev. B}\ }\textbf {\bibinfo {volume} {86}},\
  \bibinfo {pages} {024522} (\bibinfo {year} {2012})}\BibitemShut {NoStop}%
\bibitem [{\citenamefont {Prozorov}\ \emph {et~al.}(2019)\citenamefont
  {Prozorov}, \citenamefont {Ko{\'{n}}czykowski}, \citenamefont {Tanatar},
  \citenamefont {Wen}, \citenamefont {Fernandes},\ and\ \citenamefont
  {Canfield}}]{npjProzorov2019}%
  \BibitemOpen
  \bibfield  {author} {\bibinfo {author} {\bibfnamefont {R.}~\bibnamefont
  {Prozorov}}, \bibinfo {author} {\bibfnamefont {M.}~\bibnamefont
  {Ko{\'{n}}czykowski}}, \bibinfo {author} {\bibfnamefont {M.~A.}\ \bibnamefont
  {Tanatar}}, \bibinfo {author} {\bibfnamefont {H.-H.}\ \bibnamefont {Wen}},
  \bibinfo {author} {\bibfnamefont {R.~M.}\ \bibnamefont {Fernandes}},\ and\
  \bibinfo {author} {\bibfnamefont {P.~C.}\ \bibnamefont {Canfield}},\
  }\bibfield  {title} {\bibinfo {title} {Interplay between superconductivity
  and itinerant magnetism in underdoped {Ba}$_{1-x}${K}$_x${Fe}$_2${As}$_2$
  (x=0.2) probed by the response to controlled point-like disorder},\ }\href
  {https://doi.org/10.1038/s41535-019-0171-2} {\bibfield  {journal} {\bibinfo
  {journal} {npj Quantum Mater.}\ }\textbf {\bibinfo {volume} {4}},\ \bibinfo
  {pages} {34} (\bibinfo {year} {2019})}\BibitemShut {NoStop}%
\bibitem [{\citenamefont {Prozorov}\ \emph {et~al.}(2014)\citenamefont
  {Prozorov}, \citenamefont {Ko\'{n}czykowski}, \citenamefont {Tanatar},
  \citenamefont {Thaler}, \citenamefont {Bud'ko}, \citenamefont {Canfield},
  \citenamefont {Mishra},\ and\ \citenamefont {Hirschfeld}}]{ProzorovPhysRevX}%
  \BibitemOpen
  \bibfield  {author} {\bibinfo {author} {\bibfnamefont {R.}~\bibnamefont
  {Prozorov}}, \bibinfo {author} {\bibfnamefont {M.}~\bibnamefont
  {Ko\'{n}czykowski}}, \bibinfo {author} {\bibfnamefont {M.~A.}\ \bibnamefont
  {Tanatar}}, \bibinfo {author} {\bibfnamefont {A.}~\bibnamefont {Thaler}},
  \bibinfo {author} {\bibfnamefont {S.~L.}\ \bibnamefont {Bud'ko}}, \bibinfo
  {author} {\bibfnamefont {P.~C.}\ \bibnamefont {Canfield}}, \bibinfo {author}
  {\bibfnamefont {V.}~\bibnamefont {Mishra}},\ and\ \bibinfo {author}
  {\bibfnamefont {P.~J.}\ \bibnamefont {Hirschfeld}},\ }\bibfield  {title}
  {\bibinfo {title} {{Effect of Electron Irradiation on Superconductivity in
  Single Crystals of {Ba(Fe}$_{1-x}${Ru}$_x$)$_{2}${As}$_{2}$ ($x = 0.24$)}},\
  }\href {https://doi.org/10.1103/PhysRevX.4.041032} {\bibfield  {journal}
  {\bibinfo  {journal} {Phys. Rev. X}\ }\textbf {\bibinfo {volume} {4}},\
  \bibinfo {pages} {041032} (\bibinfo {year} {2014})}\BibitemShut {NoStop}%
\bibitem [{\citenamefont {Damask}\ and\ \citenamefont
  {Dienes}(1963)}]{Damask1963}%
  \BibitemOpen
  \bibfield  {author} {\bibinfo {author} {\bibfnamefont {A.~C.}\ \bibnamefont
  {Damask}}\ and\ \bibinfo {author} {\bibfnamefont {G.~J.}\ \bibnamefont
  {Dienes}},\ }\href@noop {} {\emph {\bibinfo {title} {{Point Defects in
  Metals}}}}\ (\bibinfo  {publisher} {Gordon \& Breach Science Publishers
  Ltd},\ \bibinfo {year} {1963})\BibitemShut {NoStop}%
\bibitem [{\citenamefont {Thompson}(1969)}]{Thompson1969}%
  \BibitemOpen
  \bibfield  {author} {\bibinfo {author} {\bibfnamefont {M.~W.}\ \bibnamefont
  {Thompson}},\ }\href@noop {} {\emph {\bibinfo {title} {{D}efects and
  {R}adiation {D}amage in {M}etals}}},\ \bibinfo {edition} {revised september
  27, 1974}\ ed.,\ Cambridge Monographs on Physics\ (\bibinfo  {publisher}
  {Cambridge University Press},\ \bibinfo {year} {1969})\BibitemShut {NoStop}%
\end{thebibliography}
%

\end{document}